\documentclass[aps,prl,preprintnumbers,twocolumn,groupedaddress,footinbib]{revtex4-1}

\pdfoutput=1

\usepackage{amsmath,amssymb,mathrsfs}
\usepackage{slashed}
\usepackage{xcolor}
\usepackage[colorlinks=true,linkcolor=blue,citecolor=blue,urlcolor=violet]{hyperref}
\usepackage{soul}
\usepackage{tikzfeynman}
\usepackage{ulem}

\newcommand{\GeV}{\mathrm{GeV}}

\newcommand{\be}{\begin{equation}}
\newcommand{\ee}{\end{equation}}
\newcommand{\Mpc}{\mathrm{Mpc}}

\begin{document}

\title{The meso-inflationary QCD axion} 

\author{Michele Redi}
\email{michele.redi@fi.infn.it}

\author{Andrea Tesi}
\email{andrea.tesi@fi.infn.it}
\affiliation{
INFN Sezione di Firenze, Via G. Sansone 1, I-50019 Sesto Fiorentino, Italy
}
\affiliation{Department of Physics and Astronomy, University of Florence, Italy
}

\begin{abstract}
We study the possibility that the axion Peccei-Quinn symmetry is spontaneously broken after the beginning of inflation.
This scenario interpolates between pre-inflationary and post-inflationary axion cosmology with significant phenomenological differences from both.
Since the axion is not present at the early stages of inflation large inflationary fluctuations are produced only at scales not constrained by CMB, avoiding the strongest isocurvature constraints. 
The energy density in isocurvature perturbations at short scales however can be comparable with the adiabatic contribution from misalignment. 
These large overdensities  can lead to the formation of axion mini-clusters and also to constraints from Lyman-$\alpha$ forest and future CMB spectral distortion measurements. If Peccei-Quinn symmetry is broken during the first O(25) e-foldings of inflation no axions are produced from the string network but contributions from the
annihilation of domain walls can further boost the abundance. This scenario is minimally realized if the Hubble scale during inflation drops below the Peccei-Quinn symmetry breaking scale but other realizations are possible.
\end{abstract}
\maketitle

\paragraph{\bf Introduction.}
The success of the QCD axion \cite{assionePQ,weinberg,wilczek}, a CP-odd real scalar field $a(x)$, stems from its elegant solutions to two apparently unrelated mysteries of the Standard Model (SM): the strong CP problem and the nature of Dark Matter (DM), see \cite{marsh,DiLuzio:2020wdo} for reviews. The first is solved because the axion is a Nambu-Goldstone boson of the U(1) Peccei-Quinn (PQ) global symmetry anomalous under QCD. This sole ingredient is enough to provide a potential for the axion, generated by QCD confinement, that in the minimum relaxes the $\theta$-angle to zero predicting the mass
\be
\frac{\alpha_s}{8\pi}\frac{a -f_a \theta}{f_a}G^A_{\mu\nu}\tilde G^{A,\mu\nu} \to m_a\approx 5.7\mu\mathrm{eV}\left(\frac{10^{12}\GeV}{f_a}\right) \,.
\ee
DM is explained by the same ingredients: at temperatures around GeV the axion starts to oscillate around the minimum of the potential behaving as cold DM. 

Such a remarkable extension of the SM however suffers from a few complications associated to ultra-violet physics. The so-called quality problem of the QCD axion can be summarized by saying that the global PQ symmetry has to be exact with exquisite precision \cite{Kamionkowski:1992mf}. The axion abundance on the other hand depends on unknown inflationary physics.
Traditionally axion cosmology falls in two categories depending on the size of the PQ breaking scale $f_a$ compared to Hubble during inflation and the maximal temperature during reheating $T_{\rm max}$:
\begin{itemize}
\item ${\rm Max}[H_I/2\pi, T_{\rm max}] > f_a$:
PQ  symmetry is restored during inflation or reheating. The axion emerges from the spontaneous breaking of the symmetry during reheating or radiation domination
together with a network of global strings. The relic abundance is uniquely determined by the non-linear evolution of the cosmic string network that however
is uncertain \cite{Hiramatsu:2012gg,Klaer:2017ond,Gorghetto:2020qws,Buschmann:2021sdq}. 
\item ${\rm Max}[H_I/2\pi, T_{\rm max}] <f_a$:
PQ symmetry is broken during the whole evolution of the visible universe so that the axion field exists always exist.
In this case the initial field value $a_0=\theta_0 f_a$ cannot be a priori determined but it can be fixed requiring that the axion reproduces the DM abundance.
Importantly dangerous axion isocurvature perturbations are generated that imply an upper bound on the scale of inflation $H_I\lesssim 10^8 \GeV (f_a/10^{12}\GeV)^{1/2}$.
\end{itemize}

In this letter we study the third scenario where PQ symmetry gets broken after $N_{\rm PQ}$ e-foldings from the beginning of inflation and not restored afterwards.
The attractive feature of breaking PQ symmetry during inflation is that the axion initially is not a physical degree of freedom so that large isocurvature perturbations are not generated on the scales tested  by the CMB. This scenario has however striking phenomenological features. 

After the PQ transition, the axion can be produced from quantum fluctuations during inflation with an amplitude $H_I/(2\pi)$. Contrary to the inflationary scenario the power is confined on small scales and can be a dominant component of the axion abundance. The mechanism is similar to gravitational production of \cite{Graham:2015rva}, and we compute it here for the QCD axion.  Such a power contributes as a calculable DM isocurvature component: on intermediate scales it generates DM substructure in the form of axion mini-clusters, while on cosmological scales it is suppressed though not irrelevant. We find indeed a strong constraint on our model arises from the matter power spectrum measured through  Lyman-$\alpha$ forest observation. In the most extreme case this scenario leads to late annihilation of unstable string/domain walls network increasing the axion abundance so that
DM is reproduced for $f_a$ as low as $10^9$ GeV.

While our discussion is framed for the QCD axion given its relevance, most of the key results hold in general 
for any spontaneous breaking of continuous symmetry that generates light Nambu-Goldstone bosons during inflation. 
This allows us to construct viable scenarios of inflationary produced DM avoiding isocurvature constraints.

\medskip
\paragraph{\bf Peccei-Quinn breaking during inflation.}
We focus on the KSVZ QCD axion model \cite{Kim,SVZ} where the PQ symmetry is broken by a complex scalar field $\Phi$ \footnote{As in the post-inflationary axion scenario we require domain wall number 1. This is realized introducing color triplet fermions charged under the PQ symmetry with quantum numbers equal to the $d$ or $u$ SM quarks.}.
The potential is given by 
\be
V = \lambda (|\Phi|^2-f_a^2/2)^2 + |\Phi|^2 \mathcal{O}\,, \quad \Phi=\frac{f_a+\rho}{\sqrt{2}}e^{i a/f_a}\,.
\ee
We assume that the operator $\mathcal{O}$ has a time dependent vacuum expectation value (VEV) during inflation, while
$\langle \mathcal{O}\rangle\approx 0$ during Big Bang cosmology. The simplest realization of this is when the operator depends on the inflaton field $\phi$,
for example $ \mathcal{O}=\kappa \phi^2$.

The VEV $\langle \mathcal{O}\rangle$ during inflation allows for a broader parameter space and also a more complex phase diagram for the PQ model
 compared to the scenarios reviewed in the introduction \cite{linde-axion}. Recent works exploiting the inflaton dynamics to suppress isocurvatures are \cite{Harigaya:2015hha,Kearney:2016vqw,Jeong:2013xta,Rosa:2021gbe,Bao:2022hsg}. Here we consider the possibility that $\langle \mathcal{O}\rangle$
restores the symmetry in the early stages of inflation.
Let us first consider $H_I/2\pi\lesssim f_a$ assuming that $H_I$ is roughly constant during inflation and that reheating does not restore the symmetry. 
If the VEV of $\mathcal{O}$ is initially large than the symmetry is unbroken. The critical condition for spontaneous symmetry breaking corresponds to
\be\label{eq:stable}
\langle \mathcal{O}\rangle \lesssim \lambda f_a^2\,.
\ee 
Note that before this condition is satisfied $\Phi$ is generically heavy compared to $H_I$, so it does not fluctuate during inflation.

A different possibility is that $H_I>2\pi f_a$  at the beginning of inflation but varies significantly afterwards so that PQ breaking takes place for $H_I\sim f_a$. This can
be realized in at least two ways. If the PQ field $\Phi$ is minimally coupled as typically assumed, stochastic inflationary fluctuations effectively restore the symmetry
provided that the quartic is sufficiently large \cite{Lyth:1992tw}. In the unbroken phase the complex field $\Phi$ is light but isocurvature perturbations are erased by the dynamics. Alternatively one can assume that the PQ field is conformally coupled to the curvature. This corresponds to $\mathcal{O} = -\frac16 R$, where $R$ is the Ricci scalar.
During inflation this coupling generates an effective mass $2H_I^2$ so that the symmetry is restored when $H_I^2>\lambda f_a^2/2$. Moreover for large $H_I$ 
there are no inflationary fluctuations of $\Phi$ in the unbroken phase due to  Weyl invariance of the action, see \cite{Redi:2022zkt} for more details.

Since the  axion abundance is determined by $f_a$, in this scenario the scale of inflation is linked to the axion decay constant, $H_I \sim  f_a\sim 10^{11}$ GeV
assuming that the DM abundance is reproduced. 
Breaking PQ symmetry through Hubble would require non trivial inflationary dynamics.
In slow-roll single-field models, the amplitude of curvature perturbations $\zeta$ at the CMB scale \cite{PLANCK}  implies 
$\epsilon|_{\rm CMB} \approx 10^{-6} (H_I^{\rm max}/10^{12}\,{\rm GeV})^2$. Here $\epsilon\equiv -{\dot H}/H^2$ and $H_I^{\rm max}$ the Hubble scale during the first e-foldings of visible inflation. 
From the equation above it follows that the variation of $H$ is initially very small.
It is however possible to construct scenarios where $\epsilon$ rapidly increases or changes discontinuously 
through a phase transition allowing for larger variation of $H$. We leave the model building aspects to future work.

\medskip
\paragraph{\bf DM abundance.}\,\,
We assume that PQ symmetry is spontaneously broken after the beginning of inflation, so that the axion emerges after $N_{\rm PQ}$ e-foldings of visible inflation.
This corresponds to a length scale today
\begin{equation}
d\sim \frac 1{k_{\rm PQ}}\equiv \frac 1 {H_0} \exp[-N_{\rm PQ}]\,,
\label{eq:kpq}
\end{equation} 
where $H_0$ is the  Hubble constant and $k_{\rm PQ}$ the comoving wave-number associated to the PQ breaking.
After the phase transition the field $\Phi$ rolls to the minimum of the potential with a random phase $\theta=a(x)/f_a$.
The exponential quasi de-Sitter expansion stretches the wavelengths so that the axion becomes homogeneous
over patches of macroscopic size. 
Assuming the correlation length to be of order $1/H_I$ at the phase transition this grows with the scale factor up to eq. (\ref{eq:kpq}) today.

Following standard computations in each patch the abundance due to misalignment $\theta$ is given by,
\begin{equation}\label{eq:axion-mis}
\frac {\rho_a^{\rm mis}}{s}\sim  \frac {1}{\sqrt{g_*(T_*)}}\frac{m_a f_a^2 \theta^2}{M_{\rm pl} T_*}\,,
\end{equation}
where $s=2\pi^2/45g_*T^3$ is the SM entropy density and  $T_*\approx$ GeV the temperature where the axion starts to oscillate.
This is similar to the post-inflationary scenario with the notable difference that the correlation length of the axion in our scenario is macroscopic. 
Neglecting possible contributions from the string network (see discussion below) the axion DM abundance is found by averaging over the domains.
One finds the energy fraction 
\begin{equation}\label{eq:omega-mis}
\Omega_a^{\rm mis} h^2 \approx 0.12 \left(\frac {f_a} {2\cdot 10^{11}\,{\rm GeV}}\right)^{7/6}\,.
\end{equation}
In deriving this expression one assumes that $m^2_a(T)= q\, m_a^2 (\mathrm{GeV}/T)^{\alpha}$, where $q$ is the ratio of topological susceptibilities $q=\chi(\GeV)/\chi(0)$. The numerical value corresponds to $\alpha=8$ as determined by lattice computations and instantons estimates
and $q\approx 1.46\times 10^{-7}$ matching the zero temperature contribution at 150 MeV.

The axion  is also produced by inflationary fluctuations that however are not strongly constrained. 
Modes with wave number $k> e^{N_{\rm PQ}} H_0$, exit the horizon after the PQ phase transition and therefore 
are produced with a flat power spectrum. Modes with larger wave-number are instead suppressed because the wave-length 
is already outside the horizon when the axion is born. We parametrize the power spectrum at the end of inflation,
\be\label{eq:power-exit}
\begin{split}
\Delta_a(\eta_e,k)&=\frac {k^3}{2\pi^2}\int d^3x e^{-i \vec k \cdot \vec x} \langle a(\eta_e, \vec{x})a(\eta_e,0)\rangle\,\\
&\approx \frac{H_I^2}{4\pi^2}\, \, \mathrm{min}[1, \frac{k^n}{k_{\rm PQ}^n}]\,.
\end{split}
\ee
The spectral index $n>0$ that controls the IR behaviour is determined by the fundamental mechanism at the origin of the symmetry breaking. 
For a phase transition taking place during inflation the power spectrum vanishes on distances larger than $1/H_I$ by causality. 
In Fourier space this gives a  power spectrum proportional to $k^3$ (white noise), see \cite{Caprini:2009fx}. We will thus take $n = 3$ in what follows.

\begin{figure}
\begin{center}
\includegraphics[width=\linewidth]{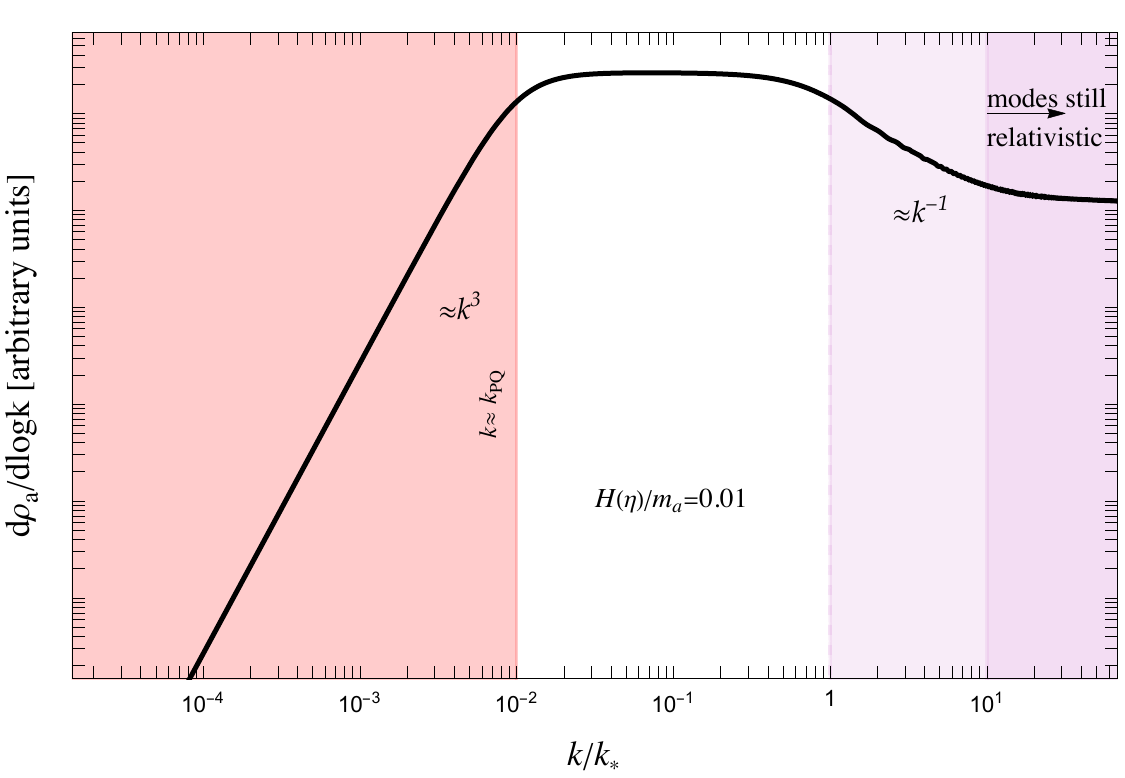}~
\caption{ \label{fig:Delta-a}\small Shape of the differential energy density $d\rho_a/d\log k$ of the axion field at a fixed time $H/m_a=0.01$. 
}
\end{center}
\end{figure}

From the power spectrum \eqref{eq:power-exit} at the end of inflation we can compute the energy density today determining its classical evolution in terms of the transfer-functions,
see for example \cite{Graham:2015rva} for analogous computations.
The energy density of the axion is at any time  $\rho_a=\frac12(g^{00}\dot a^2 - g^{ij}(\partial_i a \partial_j a)+m_a^2 a^2)$ so that
\be\label{eq:rhoX}
\frac{d\rho_a^{\rm inf}}{d\log k}= \frac{\Delta_a(\eta_e,k)}{2}\, \left[g^{00}|\dot u_k|^2-(g^{ij} k_i k_j-m_a^2)|u_k|^2\right]\,,
\ee
where $u_k$ is the transfer function, i.e. solutions to the wave-equation, from the end of inflation (with initial conditions $u_k|_{\rm exit}=1$, $\dot u_k|_{\rm exit}=0$) until the time of interest. The numerical result is found in Fig. (\ref{fig:Delta-a}) at a fixed time.
The differential energy density today is approximately given by \cite{Redi:2022zkt}
\begin{equation}\label{eq:axion-rho-inf}
\frac 1 s \frac{d \rho_a^{\rm inf}}{d \log k}\approx \frac {1}{\sqrt{g_*(T_*)}} \frac{H_I^2}{(2\pi)^2} \frac{m_a}{M_{\rm pl}T_*}  \left\{\begin{array}{lcl} 
\displaystyle   \frac{k_*}{k} & & k> k_*\,,\\
\displaystyle 1  & & k \in[k_{\rm PQ},k_*] \,,\\
\displaystyle \frac{k^3}{k_{\rm PQ}^3} & & k< k_{\rm PQ} \,,\\
\end{array}
\right.
\end{equation}
where $k_*\approx a_{\rm eq}\sqrt{m_a(T_*) H_{\rm eq}}$ and $a_{\rm eq}=1/3300$ is the scale factor at matter-radiation equality.
This expression leads to an abundance from inflationary perturbations,
\begin{equation}\label{eq:axion-inf}
\frac {\rho_a^{\rm inf}} s\approx \frac {\log(k_*/k_{\rm PQ})}{\sqrt{g_*(T_*)}} \frac{H_I^2}{(2\pi)^2} \frac{m_a}{M_{\rm pl}T_*}\,,
\end{equation}
valid for $k_{\rm PQ} < k_*$. This condition demands $H_0 e^{N_{\rm PQ}}\lesssim k_*$ and thus $N_{\rm PQ}\lesssim 25$ . For $k_{\rm PQ} > k_*$  the power in fluctuations is still significant but suppressed compared to the zero mode. In terms of the fractional energy density it corresponds to
\begin{equation}
\frac{\Omega_a^{\rm inf}}{\Omega_a^{\rm mis}}\approx \log(\frac{k_*}{k_{\rm PQ}}) \times\left( \frac{H_I}{2\pi f}\right)^2\,.
\label{eq:abmin}
\end{equation}

Let us comment on our results. For $H_I\approx 2\pi f$ the abundance from inflationary fluctuations \eqref{eq:axion-inf} is comparable to the one from misalignment \eqref{eq:axion-mis}. 
This is naturally realized in the scenario where the PQ breaking is due to the coupling to curvature.
Note that the power in isocurvature perturbations is peaked at intermediate scales $k \gtrsim k_{\rm PQ}$ not constrained by cosmological observations.

\bigskip

\medskip
\paragraph{\bf Cosmic strings and domain walls.}\,\,
The scale $k_{\rm PQ}$ in eq. (\ref{eq:kpq}) corresponds to modes that re-enter the horizon
at a temperature
\begin{equation}
T_{\rm PQ} \approx T_0 \exp[N_{\rm PQ}+4]\,,
\label{eq:TPQ}
\end{equation}
where we assumed radiation domination. 

Inflationary perturbations of the axion re-enter the horizon for $T> T_{\rm PQ}$.
On top of these the PQ phase transition itself produces inhomogeneities that will appear at somewhat smaller scales or equivalently larger temperatures,
depending on the details of the dynamics. Normally the breaking of PQ symmetry 
in the post-inflationary QCD axion generates a network of cosmic strings that quickly reaches a scaling
solution emitting axions. The final DM abundance depends on the spectrum of the axion emitted that is difficult to determine precisely \cite{Hiramatsu:2012gg,Klaer:2017ond,Gorghetto:2020qws,Buschmann:2021sdq}. 
In our scenario the formation of the string network can be vastly modified compared to the standard picture.
The string network can only form at a temperature $ T_{\rm net} \gtrsim T_{\rm PQ}$ where the field becomes strongly inhomogeneous.
Since the axions are emitted with energy of order Hubble this process is only possible if $H>m_a(T)$ at that time. It follows that the axion can only be emitted for $T_{\rm net} \gtrsim$ GeV. Conservatively we will assume that no axions are emitted if $T_{\rm PQ}< 150$ MeV. Using (\ref{eq:TPQ}) this corresponds 
to a number of e-foldings $N_{\rm PQ}<23$. 

While contributions from the string network would be suppressed (as we have implicitly assumed in eq. (\ref{eq:omega-mis})) contributions to the abundance from domain 
walls are enhanced. For $T_{\rm net}< 150$ MeV  the inhomogeneities associated to PQ breaking re-enter the horizon after the QCD phase transition.
Assuming domain wall number 1 an unstable system of topological defects emerges that quickly annihilates into axions. 
We can estimate the energy density assuming one domain wall per Hubble volume at that time, $\rho_{\rm dw}\approx \sigma H\approx 9 m_a f^2 H$ \cite{Hiramatsu:2012gg,precisely}. Since the energy density is proportional to $T^2$ the abundance $\rho_{\rm dw}/s$ grows if $T_{\rm net}$ is small and dominates the energy 
budget if $T_{\rm net}\ll$ GeV. The net effect is equivalent to a slow decay of domain walls in the post-inflationary axion scenario.

The abundance of domain walls can be written as,
\be
\Omega_a^{\rm dw} \lesssim \Omega_a^{\rm mis} \, e^{(N_{\rm dw}-N_{\rm net})}\,, ~~~~~N_{\rm net}< N_{\rm dw}\,,
\ee
where the exact value of $N_{\rm dw}$ depends on the details of the phase transition and the spectrum of axions produced by the annihilation 
of the network. We estimate $20\lesssim N_{\rm dw}\lesssim 23$.  Note that the energy density above corresponds to an isocurvature component with a power spectrum peaked at scales of order $k_{\rm PQ}$ or slightly larger. For $N_{\rm net}< N_{\rm dw}$, the axion abundance is dominated by the production from domain walls and the cosmological relic density is reproduced for  $f_a\lesssim 10^{11}$ GeV, up to the astrophysical bound $f_a\gtrsim10^8$ GeV. This implies $N_{\rm PQ}\gtrsim 20$.
Scenarios with domain wall number greater than 1 are excluded as in the  post-inflationary QCD axion scenario.

\medskip
\paragraph{\bf Inflationary isocurvature power spectrum.}\,\,
We now discuss one of the main phenomenological consequence of our scenario: the power spectrum of axion DM over-densities induced by inflationary fluctuations.
The presence of the population \eqref{eq:axion-inf} modifies the QCD axion DM over-densities at small scales introducing an iso-curvature component. We can compute the power-spectrum $\Delta_{\delta a}^{\rm iso}(k,\eta)$ associated to isocurvature over-densities $\delta_a\equiv \delta\rho_a/\langle\rho_a\rangle$ as
\be
\Delta_{\delta_a}^{\rm iso}(\eta,k) =\frac {k^3}{2\pi^2}\int d^3x e^{-i \vec k \cdot \vec x} \langle \frac{\delta\rho_a(\eta, \vec x)}{\bar \rho_a} \frac{\delta\rho_a(\eta,0)}{\bar \rho_a}\rangle \bigg|_{\rm iso}
\ee
where the homogeneous $\bar \rho_a$ contains both the misalignment and inflationary contributions. Here $\delta\rho_a(\eta,\vec x)$ is computed only with the contributions from eq.~\eqref{eq:axion-inf}. On the relevant scales, after $H(T)=m_a(T)$, the energy density is simply given by $\rho_a(\eta,\vec x)\approx m_a^2 a^2(\eta,\vec x)$, so that fluctuations in the above equation can be computed directly from the 2-point function $\langle a(x)a(0)\rangle$ and hence from the power spectrum $\Delta_a(\eta,k)$. By means of this approximation we obtain
\be
\langle \frac{\delta\rho_a}{\bar \rho_a} \frac{\delta\rho_a}{\bar \rho_a}\rangle \bigg|_{\rm iso}\approx \frac{\langle a(\eta,x)a(\eta,0)\rangle^2+ 4 a_0^2 \langle a(\eta,x)a(\eta,0)\rangle}{(a_0^2 + \langle a(\eta)^2\rangle)^2}\,.
\ee
This expression shows that for the relevant modes $k_*>k>k_{\rm PQ}$ there could be sizable over-densities if the misalignment and inflationary populations have comparable sizes.  From the explicit form of the correlation of over-densities we also see that the density contrast $\delta_a(x)$ has some degree of non-gaussianity, since only the fundamental field $a$ is a gaussian variable. 
The power spectrum can be cast in the form 
\begin{eqnarray}\label{eq:power-iso-full}
\Delta_{\delta_a}^{\rm iso}(\eta,k)&=& \frac{\Omega_{a,\rm inf}^2}{\Omega_a^2 \langle a^2 \rangle^2} \frac{k^3}{4 \pi }\int d^3q  \frac{\Delta_a(\eta,q)\Delta_a(\eta,|\vec k -\vec q|)}{q^3 |\vec k -\vec q|^3} \,\nonumber\\
 &+&\, 4 \frac{\Omega_{a,\rm inf} \Omega_{a,\rm mis}}{\Omega_a^2 \langle a^2 \rangle}\Delta_a(k)\,,
\end{eqnarray}
where $\Delta_a$ is to be interpreted as averaged upon oscillations.
This is the power spectrum as a function of time, to be computed at the epoch of interest depending on the observables under consideration. We notice that the isocurvature power spectrum depends on two contributions. The first line of eq.~\eqref{eq:power-iso-full} shows the pure quantum (non-gaussian) contribution proportional to the square of the axion power spectrum (a situation also found in vector DM \cite{Graham:2015rva}), while the second line is the usual iso-curvature power spectrum in presence of misalignment and it is gaussian.

\begin{figure}
\begin{center}
\includegraphics[width=\linewidth]{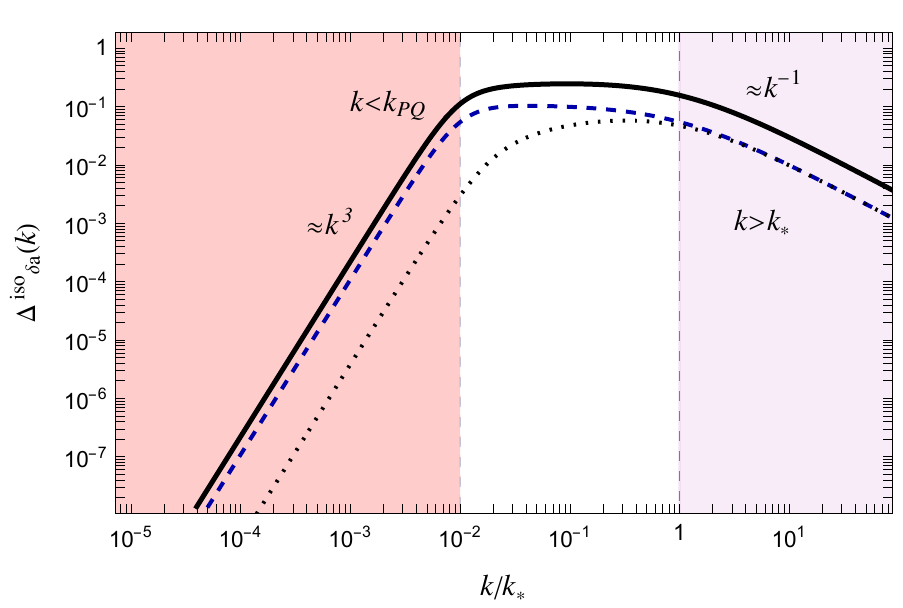}~\\
\caption{ \label{fig:iso}\small Isocurvature power spectrum $\Delta_{\delta_a}^{\rm iso}$ for  $\Omega_{a}^{\rm inf}=\Omega_{a}^{\rm mis.}$. We assume no contribution 
to the abundance from string-domain wall network. Solid black line full prediction (\ref{eq:power-iso-full});  blue dashed line contribution linear in the primordial power spectrum; dotted line quadratic contribution.}
\end{center}
\end{figure}

The amplitude $\Delta_{\delta_a}^{ \rm iso}(k,\eta)$ computed numerically (averaging out the oscillations) is found in figure \ref{fig:iso}. 
One can show that irrespectively of the shape of $\Delta_a(\eta,k)$ for $k<k_{\rm PQ}$ the contribution from the first line goes as $k^3$ at small momenta \footnote{
With a change of variable eq.~\eqref{eq:power-iso-full} can be written as 
$$\tiny
\Delta_{\delta_a}^{\rm iso}= \frac{\Omega_{a,\rm inf}^2}{\Omega_a^2 } \frac{k^2}{2  }\int \frac{dq}{q^2} \frac{\Delta_a(q)}{\langle a^2\rangle}\int_{|q-k|}^{q+k} \frac{dp}{p^2}\frac{\Delta_a(p)}{\langle a^2\rangle}+\cdots\,.
$$
}.
The pure quantum contribution (dotted line in figure \ref{fig:iso}) scales as 
\be\label{eq:power-iso-cosmo}
\Delta_{\delta_a}^{\rm iso}(\eta,k)\approx \frac{\Omega_{a,\rm inf}^2}{\Omega_a^2}\frac{k^3}{3\log^2(k_*/k_{\rm PQ})k_{\rm PQ}^3}, \quad k\ll k_{\rm PQ},
\ee
which is then power suppressed at cosmologically large scales. On the contrary, on shorter scales $k\gtrsim k_{\rm PQ}$ there is order one power that could lead to effects in structure similar to the ones studied in \cite{Gorghetto:2022sue}.

\medskip
\paragraph{\bf Bounds from CMB and large scale structure.}
The power spectrum \eqref{eq:power-iso-full}, evaluated at a cosmologically relevant time, corresponds to an isocurvature perturbation of DM. 
The infrared tail $k^3$ of the spectrum, despite being power suppressed compared to the peak, can be sizable compared to the DM adiabatic perturbations
leading to constraints that we now discuss.

There are two types of constraints that one can infer from cosmological scales: the photon power spectrum measured by PLANCK \cite{PLANCK} and the matter power spectrum tested by galaxy surveys and Lyman-$\alpha$ forest data. The first corresponds to a modification of the photon density perturbations in presence of DM isocurvature component \cite{Hu:1995en}. This tests the power spectrum on the largest scales corresponding to $k\sim 0.05\, {\rm Mpc}^{-1}$ (see also the study in \cite{Feix:2020txt}).
The second constraint is due to the fact that when DM isocurvature perturbations re-enter the horizon they start behaving as ordinary density perturbations, see \cite{Beltran:2005gr}. While the CMB provides the most precise measurement, in light of the growth of the amplitude (\ref{eq:power-iso-cosmo}) with $k^3$, the strongest constraints actually arises from the matter  power spectrum measured at  the shorted scales. Therefore, while on CMB scales we use bounds from \cite{PLANCK}, on shorter scales the effect on the matter power spectrum is equivalent to a rescaling of the adiabatic curvature perturbations. We have checked this numerically using the code \texttt{CLASS} \cite{CLASS}.

\begin{figure}
\begin{center}
\includegraphics[width=\linewidth]{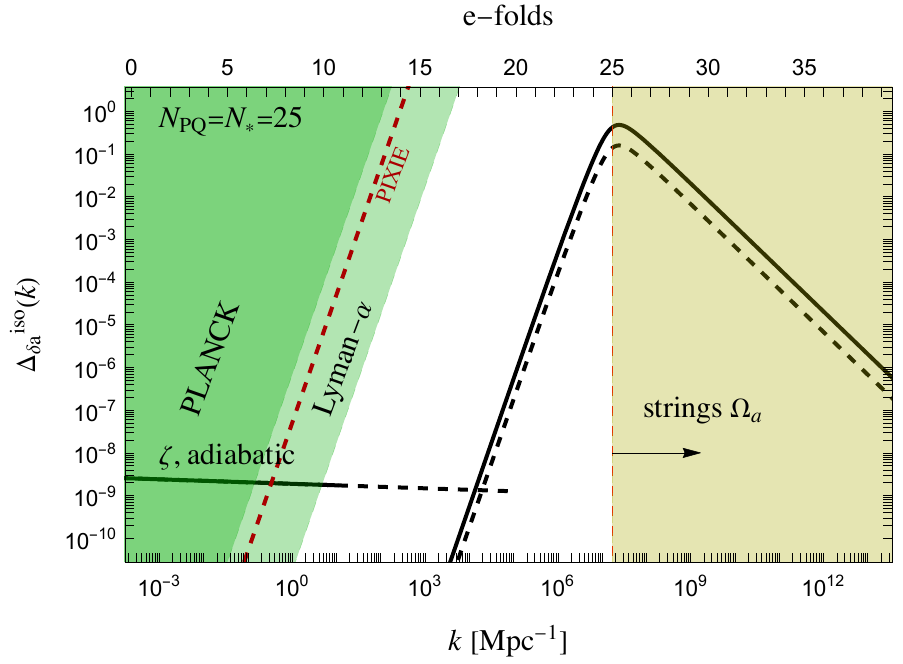}~\\
\caption{ \label{fig:iso-bound}\small Power spectrum of axion DM isocurvatures as a function of $k$ and the number of e-folds of inflation. $\Omega_{a}^{\rm inf.}=\Omega_{a}^{\rm mis.}$ (solid line), $\Omega_{a}^{\rm inf.}=0.1\Omega_{a}^{\rm mis.}$ (dashed line) for $N_{\rm PQ}=25$. Green regions would be excluded from CMB and Lyman-$\alpha$ cosmological data. In the light yellow region axion strings form at $T>$ GeV. Also shown the primordial adiabatic power spectrum of  curvature perturbations $\zeta$.}
\end{center}
\end{figure}

To determine the bounds we parametrize the isocurvature power spectrum as $\Delta_{\delta_a}^{\rm iso}(k)=A_{\rm iso} (k/k_0)^3$, where $k_0=0.05\,\Mpc^{-1}$ is the CMB pivot scale. Observations at different scales will set different bounds on $A_{\rm iso}$. From PLANCK \cite{PLANCK} we derive $A_{\rm iso}|_{\rm CMB}\lesssim 0.8\times 10^{-10}$. For smaller scales, where Lyman-$\alpha$ data are important, we find conservatively that $A_{\rm iso}|_{\rm Lyman-\alpha}\lesssim  10^{-14}$. This number is obtained requiring $\Delta_{\delta_a}^{\rm iso}(5\,\Mpc^{-1})$ is at most equal to amplitude of curvature pertubarbations $\Delta_\zeta\sim 10^{-9}$.
For an isocurvature amplitude proportional to $k^3$, this effect  manifests itself as a plateau in the matter power spectrum at large wave-numbers.
The bounds on the isocurvature power spectrum are shown in figure \ref{fig:iso-bound}. Using  (\ref{eq:power-iso-cosmo}) the bounds correspond to $N_{\rm PQ}>13, 16$  respectively. A detailed study will appear elsewhere.

On even shorter scales a large isocurvature perturbation produces $\mu$-distortions in the CMB. Spectral distortions \cite{SZ,Chluba:2011hw} can be sensitive to the photon power spectrum \cite{mu-inflation}. In our case the presence of O(1) axion DM isocurvature on short scales alters the power spectrum of the photons at horizon re-entry on the corresponding scales, $\Delta_\gamma|_{k\eta\approx1}$, prior to their sub-horizon evolution \cite{Hu:1995en}.  The large primordial contrast $\delta_{a,\rm iso}$ gives approximately \cite{Hu:1995en},
$\Delta_\gamma|_{k\approx a H} \approx O(10) \, k_{\rm eq}^2/k^2\, \Delta_{{\delta_a}}^{ \rm iso}$, which allows us to estimate the $\mu$-distortion as $\mu\approx \Delta_\gamma(\Mpc^{-1})$. Due to the suppression $k_{\rm eq}^2/k^2\approx 10^{-4}$ for the relevant modes, present COBE and FIRAS data \cite{COBE} do not give competitive constraints. This is due to the fact that DM density fluctuations are suppressed in radiation domination. Future experiments such as PIXIE \cite{PIXIE} will set a bound  $A_{\rm iso}\lesssim 10^{-11}$ \cite{iso-mu}.  In our model this corresponds to $N_{\rm PQ}\gtrsim 14$ shown as a dashed red line in figure \ref{fig:iso-bound}.

\medskip
\paragraph{\bf Miniclusters and collapsed objects.}\,\, In ~\cite{Gorghetto:2022sue} 
it was considered the impact of an O(1) peak at short scales in the power spectrum of inflationary produced Stueckelberg dark photons. 
Similarly to the axion scenario the isocurvature spectrum is maximal at $k_* \sim a_{\rm eq} \sqrt{H_{\rm eq} M}$ \cite{Graham:2015rva}.
It was shown that for $k\sim k_*$  quantum pressure is marginally irrelevant and DM can collapse into bound (solitonic) structures 
soon after matter-radiation equality, owing to perturbations being almost non-linear at horizon re-entry.  
While we foresee analogous results for our QCD axion, we also would like to highlight qualitatively different behaviour. 
Differently from the  Stueckelberg dark photon, the scale $k_*$ is smaller because it is controlled by $m_a(T_*)$ where the axion starts to oscillate.
This leads to larger structures compared to dark photon scenario where the quantum pressure is also suppressed. 
Moreover the QCD axion displays a large power on all scales between $k_{\rm PQ}\lesssim k\lesssim k_*$.  

One should also consider the contribution to the power spectrum from the overdensities at the QCD phase transition leading to axion mini-clusters \cite{Kolb:1993zz}.
For $k_{\rm PQ}\lesssim 23$ the abundance can be dominated by the delayed annihilation of the string-domain wall network. In this case the peak of
isocurvature arises at smaller wave numbers producing larger mini-clusters than in the standard post-inflationary scenario with size up to $10^{10}$ km.

\medskip
\paragraph{\bf Conclusions.}\,
We presented a novel axion scenario where the Peccei-Quinn symmetry of the QCD axion is broken after the beginning of inflation.
This scenario interpolates between  pre-inflationary and post-inflationary axion DM with new phenomenological features. 
Differently from traditional scenarios, inflation produces large isocurvature perturbations but these are not excluded
by CMB data because they arise at scales much smaller than Mpc. The isocurvature component  is sizable and can even
dominate the abundance if the phase transition takes place between 20 and 25 e-foldings from the beginning of inflation.
At the lower end the collapse of the long-lived domain wall network boosts the axion abundance requiring smaller $f_a$ to reproduce the cosmological abundance, 
eventually in conflict with stellar cooling bounds. In this limit we predict larger axion miniclusters than in the standard post-inflationary axion scenario.

Due to the large quantum inhomogeneities, axion solitonic structures might be formed with size around $10^{7}-10^{10}$ km.
Interestingly the infrared tail of the isocurvature power spectrum is constrained through the matter power spectrum measured
by the Lyman$-\alpha$ forest. Future determination of the isocurvature power spectrum at short scales might thus test the phase transitions that produced the QCD axion.

In this work we focused  on the QCD axion but the same mechanism works in general for inflationary production of DM,
avoiding isocurvature constraints. In particular this can be applied to axion-like particles and more in general to light scalar DM,
by studying the phase transition associated to the generation of a light particle during inflation. 

\medskip
\paragraph{ \it \small \bf Acknowledgements.}\,\,
{\small This work is supported by MIUR grants PRIN 2017FMJFMW and 2017L5W2PT. 
We would like to thank Alessio Notari for valuable discussions.}


\bibliography{axion_biblio}

\end{document}